\documentclass[twocolumn,aps,pra,10pt,showpacs]{revtex4-1}
\usepackage{bm}
\usepackage{amsfonts}
\usepackage{amssymb}
\usepackage{amsmath}
\usepackage{graphicx}

\begin{document}

\title{Polarization-dependent heating of the cosmic microwave background radiation by a magnetic field}
\author{Zofia Bialynicka-Birula}\affiliation{Institute of Physics, Polish Academy of Sciences\\
Aleja Lotnik\'ow 32/46, 02-668 Warsaw, Poland}
\author{Iwo Bialynicki-Birula}\email{birula@cft.edu.pl}
\affiliation{Center for Theoretical Physics, Polish Academy of Sciences\\
Aleja Lotnik\'ow 32/46, 02-668 Warsaw, Poland}

\begin{abstract}
The changes in the cosmic microwave background (CMB) spectrum seen as an increase of temperature due to a strong magnetic field are determined and their influence on the polarization of the radiation is exhibited. The effect is due to the coupling of the CMB photons to the magnetic field in the QED vacuum via the interaction with virtual pairs. In spite of the fact that the distortion of the CMB spectrum for magnetic fields that exist in the vicinity of magnetars is quite large, this effect is very difficult to detect at present because the required angular resolutions is not yet available.
\end{abstract}
\maketitle

\section{Introduction}

The most fascinating explanation of recent observations of the polarization of the CMB radiation \cite{bicep} is the hypothesis of primordial gravitational waves. However, there are other sources of polarization-dependent deformations of the CMB spectrum that have been pointed out (galactic dust \cite{ms} and topological defects \cite{mp,liz}).

In this paper we explore a similar effect: the polarization-dependent distortion of the CMB spectrum in the presence of magnetic fields. The coupling of the magnetic field to the CMB is due to the field-dependent QED vacuum polarization. In order to determine precisely this effect we quantize electromagnetic waves propagating on the magnetic background. Then we consider the properties of the thermal state of the quantized radiation. In our analysis we shall use the results obtained previously in the classical theory \cite{bb,sa}.

\section{Canonical formulation of the field dynamics}

Our starting point is a relativistic Lagrangian $L(S,P)$ where $S$ and $P$ are the scalar and the pseudoscalar invariants of the electromagnetic field,
\begin{subequations}
\begin{align}\label{inv}
S&=-\frac{1}{4}f_{\mu\nu}f^{\mu\nu}=\frac{1}{2}({\bm E}^2-c^2{\bm B}^2),\\
P&=-\frac{1}{4}f_{\mu\nu}{\hat f}^{\mu\nu}=c{\bm E}\cdot{\bm B}.
\end{align}
\end{subequations}
We will employ the Heisenberg-Euler Lagrangian $L_{HE}$ derived in QED \cite{he} under the assumption that the electric and magnetic fields {\em do not change} in space and time,
\begin{widetext}
\begin{align}\label{he}
L_{HE}=\epsilon_0\left[S+\alpha{\mathcal E}_c^2\int_0^\infty\!d\eta
\frac{ e^{-\eta}}{\eta^3}\left(1-\frac{\eta a}{\tan(\eta a)}\frac{\eta b}{\tanh(\eta b)}-\frac{\eta^2}{3}(a^2-b^2)\right)\right],
\end{align}
\end{widetext}
where $\alpha$ is the fine structure constant and ${\mathcal E}_c=m^2c^3/e\hbar=1.32\times 10^{18}$V/m is the characteristic value of the electric field built from the electron mass and the universal constants, called the critical field by Heisenberg and Euler. The corresponding magnetic critical field is ${\mathcal B}_c=4.4\times 10^9$T. The invariants $S$ and $P$ enter the Lagrangian (\ref{he}) through the expressions:
\begin{subequations}
\begin{align}\label{ab}
a&=\sqrt{\sqrt{S^2+P^2}-S}\,/{\mathcal E}_c,\\
b&=\sqrt{\sqrt{S^2+P^2}+S}\,/{\mathcal E}_c.
\end{align}
\end{subequations}
The Heisenberg-Euler Lagrangian can also be used for fields whose variation in space and time is slow on the characteristic quantum scale: $\hbar/mc=3.86\times 10^{-13}$m and $\hbar/mc^2=1.288\times 10^{-21}$s. We apply the theory based on this Lagrangian to the description of the cosmic microwave background (CMB) radiation in the presence of a magnetic field.

\begin{figure}
\vspace{0.5cm}

\includegraphics[scale=0.9]{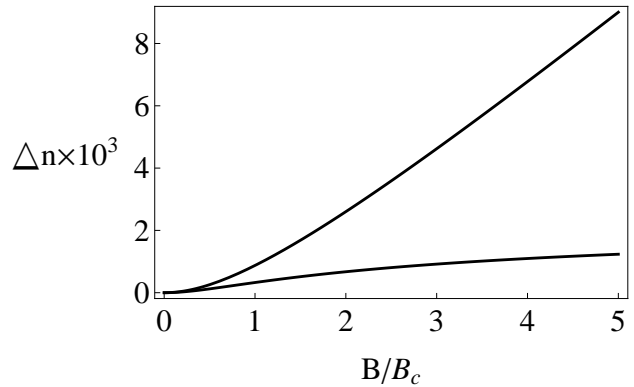}
\caption{The change in the refractive indices $\Delta n_{_\parallel}$ (lower curve) and $\Delta n_{_\perp}$ (upper curve) induced by the background magnetic field $\bm B$.}\label{fig1}
\end{figure}

\begin{figure}
\includegraphics[scale=0.9]{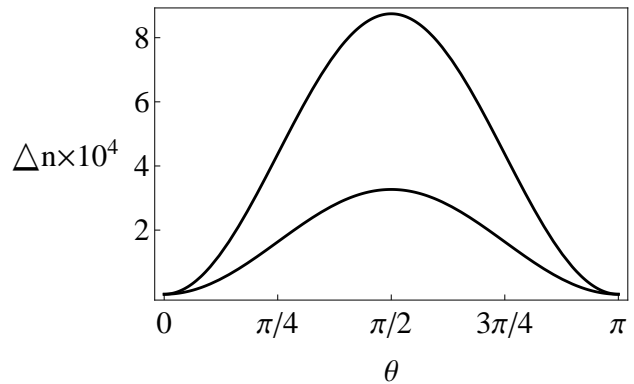}
\caption{The change in the refractive indices $\Delta n_{_\parallel}$ (lower curve) and $\Delta n_{_\perp}$ (upper curve) for the critical magnetic field as a function of the angle between the magnetic field and the direction of observation.}\label{fig2}
\end{figure}

Following the procedure introduced in Ref.~\cite{bb}, we separate the electromagnetic field into a strong, constant magnetic field ${\bm B}$ and a weak wave $({\bm e},{\bm b})$,
\begin{align}\label{dec}
S&=\frac{1}{2}\left({\bm e}^2-c^2({\bm B}+{\bm b})^2\right),\\
P&=c{\bm e}\cdot({\bm B}+{\bm b}).
\end{align}
Next, we expand the Lagrangian keeping only the quadratic terms in $({\bm e},{\bm b})$ (linear terms do not contribute to the equations of motion),
\begin{align}\label{q}
L^{(2)}=\frac{1}{2}\left[\gamma_s({\bm e}^2-c^2{\bm b}^2)+\gamma_{ss}c^2({\bm B}\cdot{\bm b})^2
+\gamma_{pp}c({\bm B}\cdot{\bm e})^2\right],
\end{align}
where
\begin{align}\label{gammas}
\gamma_s=\frac{\partial L_{HE}}{\partial S},\quad\gamma_{ss}=\frac{\partial^2 L_{HE}}{\partial S^2},\quad\gamma_{pp}=\frac{\partial^2 L_{HE}}{\partial P^2}.
\end{align}
These derivatives of the Heisenberg-Euler lagrangian are to be evaluated at $S=-c^2{\bm B}^2/2$ and $P=0$. Aiming at the quantization of the $({\bm e},{\bm b})$ fields, we find it more convenient to use not the Lagrangian but the Hamiltonian formulation. The classical Hamiltonian expressed as a function of divergenceless canonical fields ${\bm d}=\partial L^{(2)}/\partial{\bm e}$ and $\bm b$ has the form:
\begin{align}\label{ham}
H=\int\!d^3r&\Big[\frac{1}{2\gamma_s}{\bm d}({\bm r},t)\cdot\frac{1+\kappa_p(1-{\hat P})}{1+\kappa_p}\cdot{\bm d}({\bm r},t)\nonumber\\
&+\frac{\gamma_s c^2}{2}{\bm b}({\bm r},t)\cdot(1-\kappa_s{\hat P})\cdot{\bm b}({\bm r},t)\Big],
\end{align}
where ${\hat P}=({\bm B}\wedge{\bm B})/{\bm B}^2$ is the projector on the direction of the magnetic field and
\begin{align}\label{kappa}
\kappa_p=\gamma_{pp}c^2{\bm B}^2/\gamma_s,\quad
\kappa_s=\gamma_{ss}c^2{\bm B}^2/\gamma_s.
\end{align}
The canonical evolution equations for $\bm d$ and $\bm b$, obtained from the Lagrangian $L^{(2)}$, are:
\begin{subequations}
\begin{align}\label{eqs}
\partial_t{\bm d}({\bm r},t)&
=\gamma_s{\bm\nabla}\times(1-\kappa_s{\hat P}){\bm b}({\bm r},t),\\
\partial_t{\bm b}({\bm r},t)&
=-\frac{1}{\gamma_s}{\bm\nabla}\times\frac{1+\kappa_p(1-{\hat P})}{1+\kappa_p}{\bm d}({\bm r},t).
\end{align}
\end{subequations}
These equations have a complete set of solutions in the form of normalized monochromatic complex mode vectors. Each mode has the ${\bm d}$ part and the ${\bm b}$ part,
\begin{subequations}
\begin{align}\label{mode}
{\bm d}^{^\parallel}_{\bm k}({\bm r},t)& =N\sqrt{\gamma_s\omega_{_\parallel}}{\bm m}\,e^{-i\omega_{_\parallel}t+i{\bm k}\cdot{\bm r}},\\
{\bm b}^{^\parallel}_{\bm k}({\bm r},t)& =Nc|{\bm k}|\sqrt{\frac{1}{\gamma_s\omega_{_\parallel}}}({\bm m}\times{\bf n})\,e^{-i\omega_{_\parallel}t+i{\bm k}\cdot{\bm r}},\\
{\bm d}^{^\perp}_{\bm k}({\bm r},t)&=Nc|{\bm k}|\sqrt{\frac{\gamma_s}{\omega_{_\perp}}}({\bm m}\times{\bf n})\,e^{-i\omega_{_\perp}t+i{\bm k}\cdot{\bm r}},\\
{\bm b}^{^\perp}_{\bm k}({\bm r},t)& =-N\sqrt{\frac{\omega_{_\perp}}{\gamma_s}}{\bm m}\,e^{-i\omega_{_\perp}t+i{\bm k}\cdot{\bm r}},
\end{align}
\end{subequations}
where
\begin{align}\label{defs}
N&=\frac{1}{\sqrt{2(2\pi)^3}},\quad{\bm n}=\frac{{\bm k}}{|{\bm k}|},\quad{\bm m}=\frac{{\bm k}\times{\bm B}}{|{\bm k}\times{\bm B}|}.
\end{align}
Following Adler \cite{sa}, we called these modes parallel (the $\bm b$ field lies in the ${\bm B}{\bm k}$-plane) and perpendicular (the $\bm b$ field perpendicular to this plane).
The mode frequencies are given by the following formulas:
\begin{align}\label{freq}
\omega_{_\parallel}&=\frac{c|{\bm k}|}{n_{_\parallel}},\quad \omega_{_\perp}=\frac{c|{\bm k}|}{n_{_\perp}},
\end{align}
where the two refractive indices are:
\begin{subequations}
\begin{align}\label{indx}
n_{_\parallel}&=\frac{1}{\sqrt{1-\kappa_s\sin^2\theta}}=1+\Delta n_{_\parallel},\\
n_{_\perp}&=\sqrt{\frac{1+\kappa_p}{1+\kappa_p\cos^2\theta}}=1+\Delta n_{_\perp},
\end{align}
\end{subequations}
and $\theta$ is the angle between the direction of propagation ${\bm n}$ and the direction of the magnetic field. The whole information about the nature of nonlinearity is contained in two functions of the magnitude of the magnetic field, $\kappa_s$ and $\kappa_p$. For the Heisenberg-Euler Lagrangian these functions do not have a closed form; they were evaluated numerically using {\em Mathematica} \cite{math}. Note that the magnetic field has no effect on waves propagating along the field ($\theta$=0). In Fig.~\ref{fig1} we show the changes $\Delta n_{_\parallel}$ and $\Delta n_{_\perp}$ in the refractive indices as functions of the magnetic field $B$ when the effect is the largest; the direction of the field is perpendicular to the direction of observation. In Fig.~\ref{fig2} we show the variation of the refractive indices with the angle of observation for the critical magnetic field $\mathcal B_c$.

\section{Quantization of radiation in the magnetic background}

Having found a complete set of modes, we can construct field operators as the following linear combinations of the annihilation and creation operators,
\begin{subequations}
\begin{align}\label{fop}
{\hat{\bm d}}({\bm r},t)=\sqrt{\hbar}\sum_\lambda\int\!d^3k\,{\bm d}^\lambda_{\bm k}({\bm r},t)a_\lambda(\bm k)+h.c.\\
{\hat{\bm b}}({\bm r},t)=\sqrt{\hbar}\sum_\lambda\int\!d^3k\,{\bm b}^\lambda_{\bm k}({\bm r},t)a_\lambda(\bm k)+h.c.,
\end{align}
\end{subequations}
where the index $\lambda$ takes on the values $~^\parallel$ and $~^\perp$. The standard commutation relations between the annihilation and creation operators,
\begin{align}\label{cr}
[a_\lambda({\bm k}),a^{\dagger}_{\lambda'}({\bm k})']=\delta_{\lambda\lambda'}\delta({\bm k}-{\bm k}'),
\end{align}
guarantee the correct canonical commutation relations between the field operators,
\begin{align}\label{ccr}
[{\hat b}_i({\bm r},t),{\hat d}_j({\bm r},t)]=i\hbar\epsilon_{ijk}\partial_k\delta({\bm r}-{\bm r}').
\end{align}
These relations confirm that the fields $\bm d$ and $\bm b$ (and not $\bm e$ and $\bm b$) form the canonically conjugate pair of fields \cite{bb0}. Their commutation relations are universal since they do not depend on the properties of the medium (in our case the background magnetic field). The substitution of the field operators into the formula (\ref{ham}) gives the quantum energy operator,
\begin{align}\label{qham}
{\hat H}=\sum_\lambda\int\!d^3k\,\hbar\omega_\lambda\, a^{\dagger}_{\lambda}({\bm k})a_{\lambda}({\bm k}).
\end{align}

\section{Spectral properties of the CMB radiation}

In the thermal state at the temperature $T$ the density matrix ${\hat\rho}$ is an exponential function of the Hamiltonian (\ref{qham}),
\begin{align}\label{rho}
{\hat\rho}=\exp(-\beta{\hat H})/{\rm Tr}\{\exp(-\beta{\hat H})\},\quad \beta=1/k_BT.
\end{align}
In the presence of the magnetic field, the Hamiltonian (\ref{qham}) contains the modified frequencies (\ref{freq}). The average density of the CMB photons per unit volume $n_\lambda(\bm k)$ in the thermal state with the wave vector $\bm k$ and polarization $\lambda$ is:
\begin{align}\label{photdens}
<n_\lambda(\bm k)>={\rm Tr}\{{\hat\rho}a^{\dagger}_\lambda({\bm k})a_{\lambda}({\bm k})\}=\frac{1}{(2\pi)^{3}(e^{\beta\hbar\omega_\lambda}-1)}.
\end{align}
The influence of the magnetic field on the spectrum shows through the modification of the frequencies. Since both refractive indices $n_\lambda$ are greater than one, the spectrum of the CMB radiation looks like coming from regions of higher effective temperature $T_{\rm eff}=n_\lambda T$. An important characteristic feature of the spectrum is its dependence on the polarization; the effective temperature for the perpendicular polarization is larger than the effective temperature for the parallel polarization. The change in the temperature due to the change of the refractive index is quite large for strong magnetic field. Already for the critical field the increase in temperature is 2.4mK for the perpendicular mode and 0.9mK for the parallel mode.  These values exceed by far the standard deviations in the measurements of the CMB temperature.


The characteristic dependence of the observable temperature of the CMB radiation on polarization would be helpful in identifying regions where there exist very strong magnetic fields. The strongest magnetic fields in the Universe exist in the vicinity of magnetars \cite{ok}. Unfortunately, the predicted distortion of the CMB spectrum by those fields cannot be detected with the present observational means. The problem lies in the smallness of the region where the magnetic field is strong, coupled with the large distance to the nearest magnetar. One may hope that in the future the QED effects of strong magnetic fields will be accessible to observations due to improvements in angular resolution and in the precision of the measurements of temperature fluctuations in CMB. At present one may hope to use the calculated effects of magnetic fields on the polarization only when the extension of regions with strong magnetic fields is much larger than that in the vicinity of magnetars.

\acknowledgments

We thank the anonymous referee for pointing out the difficulties in observing the effects predicted in this paper. This research was financed by the Polish National Science Center Grant No. 2012/07/B/ST1/03347.

\end{document}